# *Fast and broadband photoresponse of few-layer black phosphorus field-effect transistors.*

*Michele Buscema\*, Dirk J. Groenendijk, Sofya I. Blanter, Gary A. Steele, Herre S.J. van der Zant and Andres Castellanos-Gomez\*.*

Kavli Institute of Nanoscience, Delft University of Technology, Lorentzweg 1, 2628 CJ Delft (The Netherlands).

ABSTRACT

Few-layer black phosphorus, a new elemental 2D material recently isolated by mechanical exfoliation, is a high-mobility layered semiconductor with a direct bandgap that is predicted to strongly depend on the number of layers, from 0.35 eV (bulk) to 2.0 eV (single-layer). Therefore, black phosphorus is an appealing candidate for tunable photodetection from the visible to the infrared part of the spectrum. We study the photoresponse of field-effect transistors (FETs) made of few-layer black phosphorus (3 nm to 8 nm thick), as a function of excitation wavelength, power and frequency. In the dark state, the black phosphorus FETs can be tuned both in hole and electron doping regimes allowing for ambipolar operation. We measure mobilities in the order of 100 cm$^2$/V s and current ON/OFF ratio larger than 10$^3$. Upon illumination, the black phosphorus transistors show response to excitation wavelengths from the visible up to 940 nm and rise time of about 1 ms, demonstrating broadband and fast detection. The responsivity reaches 4.8 mA/W and it could be drastically enhanced by engineering a detector based on a PN junction. The ambipolar behavior coupled to the fast and broadband photodetection make few-layer black phosphorus a promising 2D material for photodetection across the visible and near-infrared part of the electromagnetic spectrum.

MAIN TEXT

Since the discovery of graphene,[1, 2] the research effort on 2D materials has rapidly increased, driven by their extraordinary electrical and mechanical properties.[3, 4] In fact, graphene, a single layer of carbon atoms arranged in a honeycomb lattice, has shown extremely high mobilities and impressively high breaking strength.[5, 6] However, the absence of a bandgap in graphene reduces its applicability in semiconducting and optoelectronic devices. This has triggered the research interest in other 2D materials with an intrinsic bandgap.[3] Silicene, a 2D layer of silicon atoms, has been considered as semiconducting counterpart of graphene but its environmental stability represents a key issue for further studies and its applicability.[7] Among the studied 2D semiconductors, the transition metal dichalcogenides (TMDCs) show promising properties for optoelectronic applications due to their direct bandgap and strong absorption.[8-12] However, these materials have demonstrated relatively slow photoresponse and, because of the large bandgap of Mo- and W-based compounds, they are suited for applications in only part of the visible range of the electromagnetic spectrum. A material with a direct and small bandgap together with fast photoresponse is needed to extend the detection range accessible with 2D materials.

Recently, a new elemental 2D material has been isolated by mechanical cleavage of black phosphorus crystals, a layered allotrope of the element phosphorus.[13-16] In bulk, black phosphorus is a semiconductor with a direct bandgap of 0.35 eV and mobilities in the order of 10000 cm$^2$/Vs.[17] Exfoliated black phosphorus flakes have been recently used as channel material in field-effect transistors (FETs) showing mobility values up to 1000 cm$^2$/V s.[13] Moreover, the small and direct bandgap opens the door to application of black phosphorus for broadband photodetection where TMDCs are limited because of their large bandgap.

In this Letter, we fabricate few-layer black phosphorus (b-P) field-effect transistors and study their optoelectronic properties as photodetectors. In the dark state, the b-P FETs present ambipolar behavior with mobilities in the order of 100 cm$^2$/Vs and current ON/OFF ratios larger than 10$^3$ for holes and 0.5 cm$^2$/Vs and ON/OFF ratios of 10 for electrons. Upon illumination, the b-P FETs show photoresponse reaching 4.8 mA/W in the visible and extending up to wavelengths in the near-infrared (940 nm). Furthermore, the b-P photodetector shows time response of 1ms (rise) and 4 ms (fall). These properties make few-layer black phosphorus a promising active 2D material in broadband and fast photodetectors across the visible and near-infrared.

We isolate few-layer b-P on a SiO$_2$ (285 nm)/Si substrate via a modified version of the mechanical exfoliation method.[18] We fabricate b-P based FET$_S$ by standard e-beam lithography, metal deposition (Ti/Au 5/50nm) and lift-off. In Figure 1a we show an optical micrograph of the fabricated b-P FET. The metallic leads were patterned after selection of thin regions in the flake by optical contrast.[19] The topography of the fabricated devices is studied by atomic force microscopy (Figure 1b). The measured height of the flake shown in Figure 1b is 8 nm, corresponding to 15 layers of b-P. Figure 1c shows a Raman spectrum of the flake in which the three peaks characteristic for pristine black phosphorus can clearly be identified.[20] We performed electrical characterization in dark and under light excitation in a probe station at room temperature and in vacuum ($10^{-5}$ mbar). Figure 1d shows a schematic of the b-P FET and the measurement circuit.

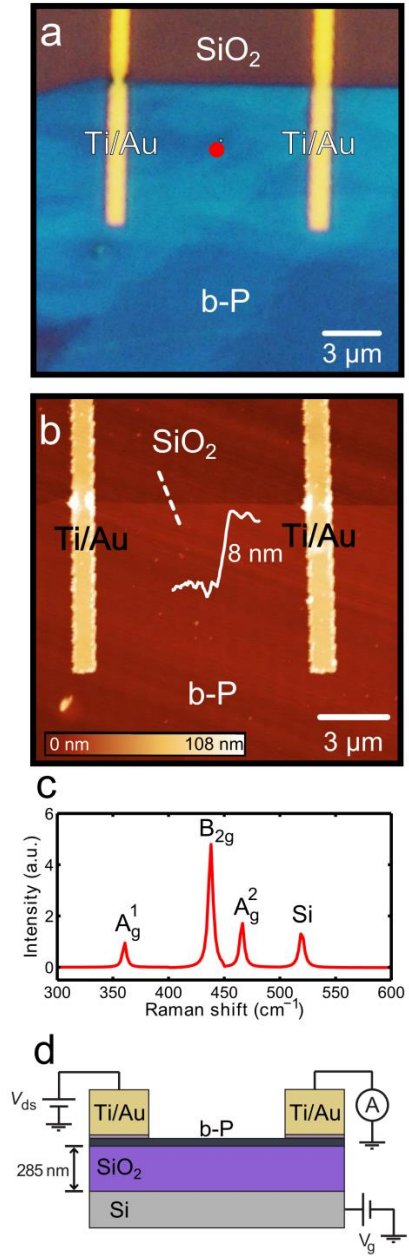

**Figure 1.** (a) Optical image of the black-phosphorus (b-P) based FET. The flake is located on a $SiO_2$/Si substrate and is contacted by Au electrodes with Ti adhesion layer. (b) Topography of the b-P based FET presented in panel (a) measured with AFM. (c) Raman spectrum of the few-layer b-P taken at the position of the red dot in panel (a). (d) Schematics of the b-P based FET and the circuit used to perform the two terminal electrical measurements.

In Figure 2 we present the electrical characterization in dark state of the device shown in Figure 1. The six studied devices display similar behavior (we refer the reader to the Supplementary Information Material). The two-terminal transfer characteristics ($I_{ds}$-$V_g$ at different fixed $V_{ds}$) in Fig. 2a show large currents at negative gate values that decrease to a minimum value at small positive gate voltages and followed by a shallow increase as the gate is swept towards more positive voltages. The inset of Figure 2a shows the transfer characteristics in a linear scale. The measured transfer characteristics describe the typical behavior of an ambipolar transistor and demonstrate that b-P FETs can operate in both holes and electron doping regimes by simply controlling the gate electric field, unlike MoS$_2$-based transistors.[21, 22] The current minimum (flatband condition, $V_{FB}$) is reached at $V_g \sim$ +10 V, indicating that the natural doping of this b-P device is p-type and that the bands at $V_g$ = 0 V are curved upwards at the contacts, facilitating hole conduction.

In the hole doping regime, the current levels can be tuned by more than 3 orders of magnitude, displaying OFF currents around hundreds of pA and ON currents of a few hundreds of nA. Lower values of ON currents are achieved for electron injection. The difference in current levels between p- and n-type operation can be explained by the band alignment imposed by the work function of the contacts that generates a small (large) Schottky barrier for hole (electron) injection (Figure 2d). From the measured transconductance, by employing the formula $\mu_{FE} = \frac{dI_{ds}}{dV_g} \frac{L}{W} \frac{1}{C_{ox} \cdot V_{ds}}$ where $L$ and $W$ are the channel length and width respectively and $C_{ox}$ is the capacitance of the gate oxide,[23] we extract mobilities ($\mu_{FE}$) in the order of 100 cm$^2$/V s for hole transport and 0.5 cm$^2$/V s for electron transport without any cleaning or annealing after fabrication. We note that these values represent a lower bound as the measurements were performed in two-terminal configuration, thus including the contact resistance. The mobility values compare well with recent measurements on thin b-P samples. [13-16, 24]

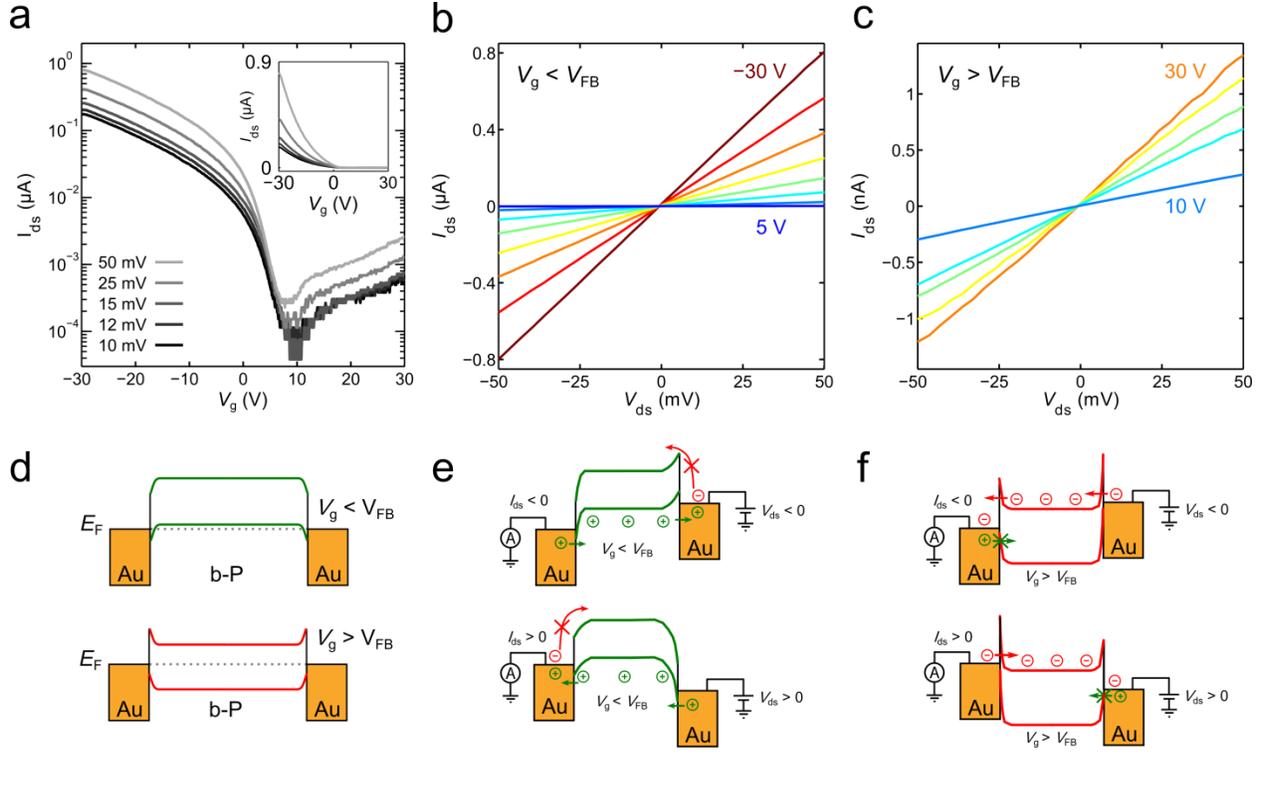

**Figure 2.** (a) Semi-log plot of the source-drain current ($I_{ds}$) versus gate voltage ($V_g$) characteristics of the b-P based FET measured at the indicated source-drain voltages. The inset shows the same measurements on a linear scale. (b) Source-drain current ($I_{ds}$) versus source-drain voltage ($V_{ds}$) characteristics measured at gate voltages below the flatband condition ($V_{FB} \sim$ 10 V) in steps of 5 V. (c) Source-drain current versus source-drain voltage characteristics measured above the flatband condition (Vg ~ 5 V) in steps of 5 V. (d) Band diagram at zero source-drain bias showing a cartoon of the Schottky barriers at the contacts for gate voltages below and above the flatband condition. $E_F$ denotes the Fermi energy. The bands are drawn for $V_g < V_{FB}$ (green lines) and for $V_g > V_{FB}$ (red line). (e) Band diagram at $V_g < V_{FB}$ for negative (top) and positive (bottom) source-drain voltage. (f) Band diagram at positive $V_g > V_{FB}$ for negative (top) and positive (bottom) source-drain voltage.

Figure 2b and Figure 2c display the measured output characteristics ($I_{ds}$-$V_{ds}$ at fixed $V_g$) for hole and electron doping, respectively. The $I_{ds}$-$V_{ds}$ curves are linear and symmetric in both cases, indicating that the Ti/Au electrodes make ohmic contact to the b-P flake. In the p-doped region ($V_g < V_{FB}$), a large current flows when applying a small bias voltage. This can be explained by having a small and thin Schottky barrier at the metal/b-P interface (Figure 2e). For electron doping ($V_g > V_{FB}$), only a small current flows through the device under bias, indicating a larger contact resistance (Figure 2f). We note that the transfer characteristics are shifted towards more positive $V_g$ values by either performing consecutive gate sweeps (Figure S2a) or applying a constant gate voltage in time (Figure S2b). Moreover, the devices relax towards their pristine behavior if the gate is grounded for a long time (Figure S2b). These signatures suggest that the gate electric field is also promoting a slowly decaying charge transfer from either the $SiO_2$/b-P interface or adsorbates on the b-P flake, as also observed in other nanodevices.[25-28] In the remainder of this Letter, we will discuss measurements performed at $V_g = 0$ V so that the gate-induced stress will not affect the measured data.

We now turn our attention to the photoresponse of the fabricated b-P FETs. The measurements are carried out by mechanically modulating the intensity of the incoming light and recording the current under constant $V_{ds} = 100$ mV and zero gate bias. Figure 3(a) shows the measured current for 940 nm, 885 nm and 640 nm excitation. The photocurrent is in phase with the excitation modulation and decreases with increasing wavelength. The clear response to excitation wavelengths up to 940 nm shows that it is still possible to photoexcite carriers over the b-P bandgap with low energy photons (1.34 eV) and extract them with a small bias voltage. Moreover, this demonstrates that b-P based devices can be operated in the near-infrared (NIR) part of the spectrum, unlike photodetectors based on other layered materials that are usually operated in the visible or ultraviolet due to their larger bandgap. [8, 10, 29-33]

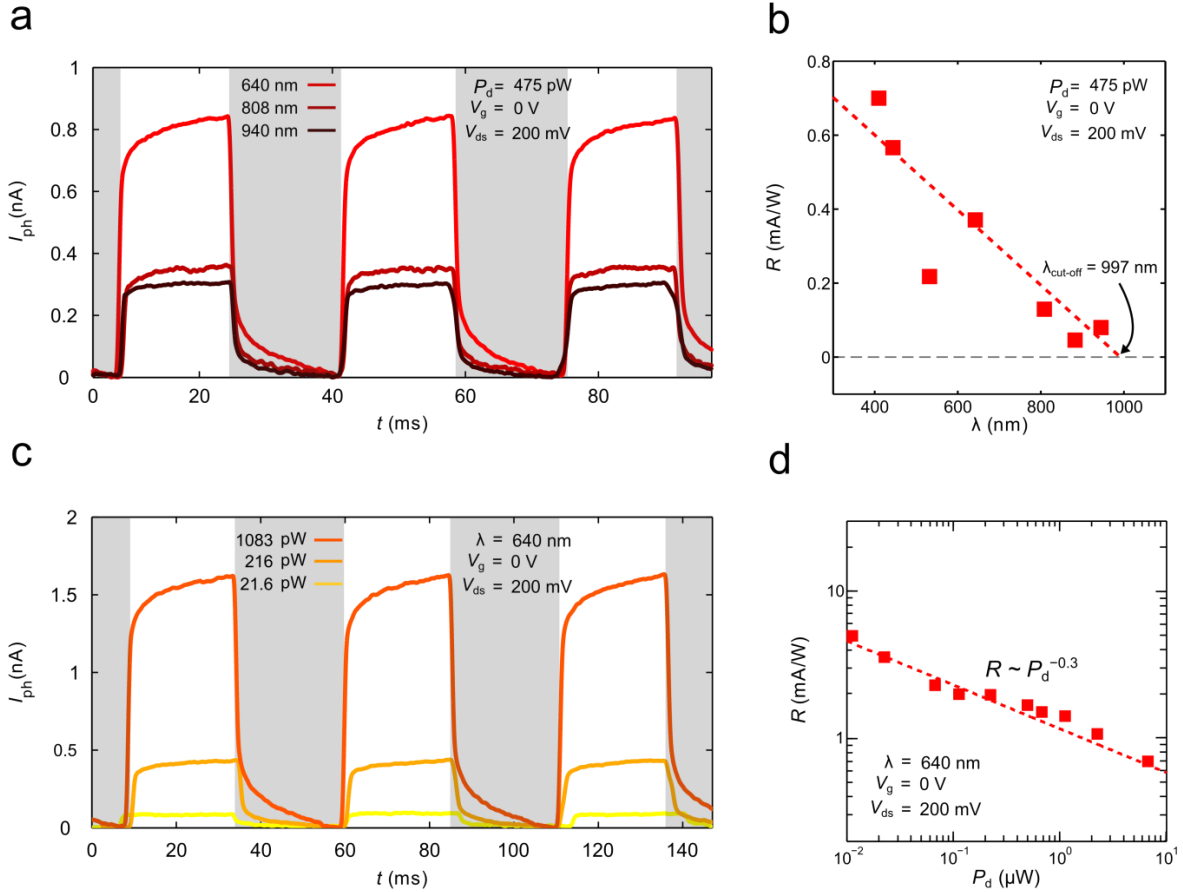

**Figure 3.** (a) Photocurrent as a function of time under modulated light excitation (~20 Hz) with different wavelengths. The shaded areas indicate the time when the light beam is blanketed by the mechanical chopper. Measurements are taken at the experimental conditions indicated in the panel. (b) Responsivity vs. excitation wavelength at constant power (red squares). The dashed red line is a linear fit to the data. (c) Photocurrent as a function of time under modulated light excitation (~20 Hz) of different powers. (d) Responsivity vs. scaled power. The red squares represent the measured values while the dashed red line is a power law fit to the data.

The responsivity ($R$) is calculated by dividing the photocurrent ($I_{ph} = I_{\text{laser ON}} - I_{\text{laser OFF}}$) by the power incident on the device area ($P_d = P_{in} \frac{A_{device}}{A_{spot}}$, where $P_{in}$ is the total optical power, $A_{device}$ is the area of the device and $A_{spot}$ is the area of the laser spot). Figure 3b shows that the responsivity increases as the excitation wavelength is decreased. The other measured devices show similar behaviour (Figure S4). Since the measurements were conducted at the same incident power for all the different wavelengths, the increase in responsivity could be due to an increase in the optical absorption of the material at shorter wavelengths. By linearly fitting the data, we can estimate the wavelength where the responsivity is zero; we find this cut-off wavelength ($\lambda_{cut-off}$) to be around 997 nm (1.24 eV). The $\lambda_{cut-off}$ value is larger than previously reported photodetectors based on $MoS_2$,[8, 29, 34] $WS_2$,[35] $GaSe$[36] and $GaS$,[32] indicating that b-P can be used as active material for broadband photodetection. Moreover, the bandgap predicted by DFT calculations for this thickness of b-P is much smaller than our estimate of the cut-off wavelength.[14] Further studies with excitation by lower energy photons should be carried out to measure the photoresponse up to the b-P optical absorption edge.

We further characterize the b-P based phototransistor by varying the excitation power at a fixed excitation wavelength (640 nm). The photocurrent increases sublinearly with increasing intensity and it reaches a few nA at high excitation powers (Figure S5). We measure responsivities ($R$) up to 4.8 mA/W with small source drain bias ($V_{ds}$ = 100 mV). These values are in the same order of magnitude of the first reports on photodetectors based on $MoS_2$ (8 mA/W with $V_{ds}$ = 1 V),[8] indicating a possibility of improvement through device engineering, as in the case of more recent $MoS_2$-based devices.[29] The measured responsivity decreases with increasing excitation power (Figure 3d). The responsivity vs. incident optical power relationship can be modeled by a power law of the following type $R \propto P_d^{\alpha-1}$ (1). By fitting the data with expression (1), we extract α ~ 0.7, indicating that the recombination kinetics of photocarriers involves both trap states and interactions between the photogenerated carriers.[37, 38] The decrease in responsivity with incident optical power is commonly observed in photodetectors based on other nanostructured materials, such as $MoS_2$,[10, 29] and colloidal quantum dots.[39, 40] This effect can be associated with a reduction of the number of photogenerated carriers available for extraction under high photon flux due to Auger processes or the saturation of recombination/trap states that influence the lifetime of the generated carriers.[38]

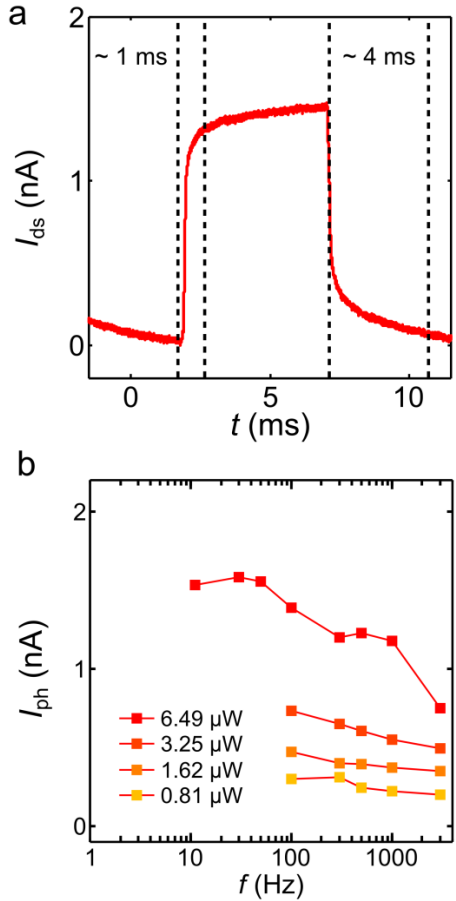

**Figure 4.** (a) Photocurrent measured in one period of modulation of the light intensity ($\lambda$ = 640 nm, $P_d$ = 6.49 µW, $V_{ds}$ = 100 mV). We can estimate a rise (fall) time of 1 ms (4 ms). (b) Photocurrent as a function of the modulation frequency for different incident power levels.

By mechanically modulating the light intensity, we measure the speed of the device response to the incoming excitation. The b-P FETs show a remarkably fast response: from one period of light excitation (Figure 4a), we measure the 10-90% rise and fall time to be 1 ms and 4 ms, respectively. From the rise time we can estimate an $f_{3dB}$ of 6.2 kHz. We measure the amplitude of the photocurrent as a function of modulation frequency and power (Figure 4b). For the highest incident power ($P_d$ = 6.49 µW), the photocurrent amplitude decreases with increasing modulation frequency and reaches the -3 dB point at 3 kHz (experimental limit), in fair agreement with the estimated $f_{3dB}$ from the rise time. Furthermore, as the incident power is reduced, the photocurrent decreases more slowly with the modulation frequency, indicating that the $f_{3dB}$ of the device is larger for lower excitation power. The reduction of $f_{3dB}$ at higher incident photon flux could be related to an increase in the rate of the recombination processes between photogenerated charge carriers, which is also consistent with the measured sublinear increase of the photocurrent with power.

In order to assess the performance of the fabricated b-P photodetector, we summarize in Table 1 the relevant figures of merit reported in literature for other devices based on semiconducting layered materials. The experimental conditions are also listed for ease of comparison. The measured responsivity of the fabricated b-P transistor is much larger than detector based on multilayer $WS_2$ and it is in the same order of magnitude as the first report of photodetection in 1L $MoS_2$. Other materials show higher responsivity values but much slower response times and limitation in the detection range. The combination of sizeable (R = 4.8 mW/A), broadband ($\lambda_{cut-off}$ ~ 997 nm) and fast ($\tau_{rise}$ ~ 1 ms) photoresponse makes b-P a very promising material for photodetection with room for improvement through further device engineering.

**Table 1.** Comparison of figures-of-merit for photodetectors based on 2D materials

| Material | Measurement conditions | Responsivity (mA/W) | Response time (ms) | Spectral range | Reference |
|---|---|---|---|---|---|
| >1L b-P | $V_{ds}$ = 0.1 V, $V_g$ = 0V, $\lambda$ = 640 nm, $P_d$ = 10 pW | 4.8 | 1 | Visible - NIR | This work |
| 1L $MoS_2$ | $V_{ds}$ = 8 V, $V_g$ = -70 V, $\lambda$ = 561 nm, P = 150 pW | 880 $10^3$ | 600 | Visible | 29 |
| 1L $MoS_2$ | $V_{ds}$ = 1 V, $V_g$ = 50 V, $\lambda$ = 532 | 8 | 50 | Visible | 8 |

| | nm, $P$ = 80 µW | | | | |
|---|---|---|---|---|---|
| > 1L MoS$_2$ | $V_{ds}$ = 1 V, $V_g$ = -2 V, $\lambda$ = 633 nm, $P$ = 50 mW/cm$^2$ | 110 | > 10$^3$ | Visible - NIR | 34 |
| > 1L WS$_2$ | $V_{ds}$ = 30 V, $V_g$ = N.A., $\lambda$ = 458 nm, $P$ = 2 mW | 21.4 10$^{-3}$ | 5.3 | Visible | 35 |
| > 1L In$_2$Se$_3$ | $V_{ds}$ = 5 V, $V_g$ = 0 V, $\lambda$ = 300 nm, $P$ = 2.08 W/m$^2$ | 395 10$^3$ | 18 | UV - NIR | 33 |
| > 1L GaTe | $V_{ds}$ = 5 V, $V_g$ = 0 V, $\lambda$ = 532 nm, $P$ = 3 10$^{-5}$ mW/cm$^2$ | 10$^7$ | 6 | Visible | 42 |
| > 1L GaSe | Vds = 5 V, Vg = 0 V, $\lambda$ = 254 nm, $P$ = 1 mW/cm$^2$ | 2.8 10$^3$ | 300 | UV - Visible | 36 |
| >1L GaS | $V_{ds}$ = 2 V, $V_g$ = 0V, $\lambda$ = 254 nm, $P$ = 0.256 mW/cm$^2$ | 4.2 10$^3$ | 30 | UV-Visible | 32 |

In conclusion, we have characterized the opto-electronic properties of b-P based FETs. In the dark state, the b-P FETs show ambipolar behavior and good mobility values in the hole doping regime. The ambipolarity opens the door to the realization of electrostatically defined PN junctions with great promise for both electronics and optoelectronics. We measure a broadband response to light excitation from the visible up to the NIR region, in contrast to photodetectors based on TMDCs and other semiconducting layered materials that are limited to the visible part of the spectrum. Our results show the potential of black phosphorus as active material for high-speed NIR detectors.

Methods

*Fabrication of black Phosphorus (b-P) based FETs:* b-P FETs were fabricated by means of an all-dry transfer technique for the b-P nanosheets, described in Refs. [18, 41] and subsequent e-beam lithography, metal deposition and lift-off. We micromechanically exfoliate b-P from a synthetic bulk crystal (99.998 %, Smart Elements) onto a viscoelastic stamp (GelFilm® by GelPak) with blue Nitto tape (Nitto Denko Co., SPV 224P). Thin b-P flakes are then identified by optical transmission and transferred onto a $SiO_2$ (285nm)/Si substrate by pressing the viscoelastic stamp surface against the $SiO_2$/Si substrate and then slowly releasing it. Both these operations are performed with a micromanipulator. Note that contact pads and optical markers are present on the surface of the $SiO_2$/Si substrate to accurately design contacts to the transferred flakes. The leads (Ti – 5 nm/Au – 50 nm) are then patterned with standard e-beam lithography (*Vistec, EBPG5000PLUS HR 100*), metal deposition (*AJA international*) and lift-off (warm acetone).

Atomic Force Microscopy (AFM) is used to determine the number of b-P layers. The AFM (*Digital Instruments D3100 AFM*) is operated in amplitude modulation mode with Silicon cantilevers (spring constant 40 N m$^{-1}$ and tip curvature <10 nm) to measure the topography and to determine the number of b-P layers.

Raman measurements are performed in a Renishaw *in via* system in backscattering configuration. The excitation is provided by visible laser light ($\lambda$ = 514 nm) through a 100× objective (NA = 0.95). To avoid laser-induced modification or ablation of the samples, all spectra were recorded at low power levels ($P_{in}$ ~ 250 μW) and after the electrical measurements.

All the opto-electronic characterization is performed in a *Lakeshore Cryogenics* probestation at room temperature under vacuum (~ 10$^{-5}$ mBar). The light excitation is provided by diode pumped solid state lasers operated in continuous wave mode (CNI Lasers). The light is coupled into a multimode optical fiber (NA = 0.23) through a parabolic mirror ($f_{reflected}$ = 25.4 mm). At the end of the optical fiber, an identical parabolic mirror collimates the light exiting the fiber. The beam is then directed into the probe station zoom lens system and then inside the cryostat. The beam spot size on the sample has a diameter of 230 ± 8 μm for all wavelengths.

ASSOCIATED CONTENT

**Supporting Information**. AFM of studied devices, effect of gate stress, transfer characteristics of the studied devices under illumination, responsivity as a function of wavelength, responsivity as a function of power, effect of constant illumination with λ = 405 nm.

AUTHOR INFORMATION

**Corresponding Author**

*Michele Buscema m.buscema@tudelft.nl

*Andres Castellanos-Gomez a.castellanos-gomez@tudelft.nl

**Author Contributions**

The manuscript was written through contributions of all authors. All authors have given approval to the final version of the manuscript

ACKNOWLEDGMENT

This work was supported by the European Union (FP7) through the program 9 RODIN and the Dutch organization for Fundamental Research on Matter (FOM). A.C-G. acknowledges financial support through the FP7-Marie Curie Project PIEF-GA-2011-300802 ('STRENGTHNANO').

References

(1) Novoselov, K.; Geim, A.; Morozov, S.; Jiang, D.; Grigorieva, M. I. K. I. V.; Dubonos, S.; Firsov, A. *Nature* **2005,** 438, (7065), 197-200.
(2) Novoselov, K.; Geim, A.; Morozov, S.; Jiang, D.; Zhang, Y.; Dubonos, S.; Grigorieva, I.; Firsov, A. *Science* **2004,** 306, (5696), 666.
(3) Novoselov, K.; Jiang, D.; Schedin, F.; Booth, T.; Khotkevich, V.; Morozov, S.; Geim, A. *Proc. Natl. Acad. Sci. U.S.A.* **2005,** 102, (30), 10451.
(4) Wang, Q. H.; Kalantar-Zadeh, K.; Kis, A.; Coleman, J. N.; Strano, M. S. *Nat. Nanotechnol.* **2012,** 7, (11), 699-712.
(5) Wang, L.; Meric, I.; Huang, P. Y.; Gao, Q.; Gao, Y.; Tran, H.; Taniguchi, T.; Watanabe, K.; Campos, L. M.; Muller, D. A.; Guo, J.; Kim, P.; Hone, J.; Shepard, K. L.; Dean, C. R. *Science* **2013,** 342, (6158), 614-617.
(6) Lee, C.; Wei, X.; Kysar, J. W.; Hone, J. *Science* **2008,** 321, (5887), 385-388.
(7) Vogt, P.; De Padova, P.; Quaresima, C.; Avila, J.; Frantzeskakis, E.; Asensio, M. C.; Resta, A.; Ealet, B.; Le Lay, G. *Phy. Rev. Lett.* **2012,** 108, (15), 155501.
(8) Yin, Z.; Li, H.; Jiang, L.; Shi, Y.; Sun, Y.; Lu, G.; Zhang, Q.; Chen, X.; Zhang, H. *ACS Nano* **2012,** 6, (1), 74-80.
(9) Shanmugam, M.; Durcan, C. A.; Yu, B. *Nanoscale* **2012,** (4), 7399-7405.
(10) Lee, H. S.; Min, S.-W.; Chang, Y.-G.; Park, M. K.; Nam, T.; Kim, H.; Kim, J. H.; Ryu, S.; Im, S. *Nano Lett.* **2012,** 12, (7), 3695-3700.
(11) Buscema, M.; Barkelid, M.; Zwiller, V.; van der Zant, H. S. J.; Steele, G. A.; Castellanos-Gomez, A. *Nano Lett.* **2013**, *13*, (2), 358–363.
(12) Fontana, M.; Deppe, T.; Boyd, A. K.; Rinzan, M.; Liu, A. Y.; Paranjape, M.; Barbara, P. *Sci. Rep.* **2013,** 3, 1634.
(13) Xia, F.; Wang, H.; Jia, Y. *arXiv:1402.0270* **2014**.
(14) Liu, H.; Neal, A. T.; Zhu, Z.; Tomanek, D.; Ye, P. D. *arXiv:1401.4133* **2014**.
(15) Qiao, J.; Kong, X.; Hu, Z.-X.; Yang, F.; Ji, W. *arXiv:1401.5045* **2014**.
(16) Li, L.; Yu, Y.; Ye, G. J.; Ge, Q.; Ou, X.; Wu, H.; Feng, D.; Chen, X. H.; Zhang, Y. *arXiv:1401.4117* **2014**.
(17) Akahama, Y.; Endo, S.; Narita, S.-i. *J. Phys. Soc. Jpn.* **1983,** 52, (6), 2148-2155.
(18) Castellanos-Gomez, A.; Buscema, M.; Molenaar, R.; Singh, V.; Janssen, L.; Zant, H. S. J. v. d.; Steele, G. A. *arXiv:1311.4829* **2013**.
(19) Castellanos-Gomez, A.; Navarro-Moratalla, E.; Mokry, G.; Quereda, J.; Pinilla-Cienfuegos, E.; Agraït, N.; Zant, H. J.; Coronado, E.; Steele, G.; Rubio-Bollinger, G. *Nano Res.* **2013,** 6, (3), 191-199.
(20) Morita, A. *Appl. Phys. A* **1986,** 39, (4), 227-242.
(21) Radisavljevic, B.; Radenovic, A.; Brivio, J.; Giacometti, V.; Kis, A. *Nature Nanotechnol.* **2011,** 6, (3), 147-150.
(22) Zhang, Y. J.; Ye, J. T.; Yomogida, Y.; Takenobu, T.; Iwasa, Y. *Nano Lett.* **2013,** 13, (7), 3023-3028.
(23) Ghatak, S.; Pal, A. N.; Ghosh, A. *ACS Nano* **2011,** 5, (10), 7707-7712.
(24) Koenig, S. P.; Doganov, R. A.; Schmidt, H.; Neto, A. H. C.; Oezyilmaz, B. *arXiv:1402.5718* **2014**.
(25) Wang, H.; Wu, Y.; Cong, C.; Shang, J.; Yu, T. *ACS Nano* **2010,** 4, (12), 7221-7228.

(26)     Cross, R. B. M.; De Souza, M. M. *IEEE Trans. Device Mater. Reliab.* **2008,** 8, (2), 277-282.
(27)     Suresh, A.; Muth, J. F. *App. Phys. Lett.* **2008,** 92, (3), 033502.
(28)     Knipp, D.; Street, R. A.; Völkel, A.; Ho, J. *J. App. Phys.* **2003,** 93, (1), 347-355.
(29)     Lopez-Sanchez, O.; Lembke, D.; Kayci, M.; Radenovic, A.; Kis, A. *Nat. Nanotechnol.* **2013,** 8, (7), 497-501.
(30)     Pospischil, A.; Furchi, M. M.; Mueller, T. *arXiv:1309.7492* **2013**.
(31)     Baugher, B. W. H.; Churchill, H. O. H.; Yang, Y.; Jarillo-Herrero, P. *arXiv:1310.0452* **2013**.
(32)     Hu, P.; Wang, L.; Yoon, M.; Zhang, J.; Feng, W.; Wang, X.; Wen, Z.; Idrobo, J. C.; Miyamoto, Y.; Geohegan, D. B.; Xiao, K. *Nano Lett.* **2013,** 13, (4), 1649-1654.
(33)     Jacobs-Gedrim, R. B.; Shanmugam, M.; Jain, N.; Durcan, C. A.; Murphy, M. T.; Murray, T. M.; Matyi, R. J.; Moore, R. L.; Yu, B. *ACS Nano* **2013,** 8, (1), 514-521.
(34)     Choi, W.; Cho, M. Y.; Konar, A.; Lee, J. H.; Cha, G.-B.; Hong, S. C.; Kim, S.; Kim, J.; Jena, D.; Joo, J.; Kim, S. *Adv. Mater.* **2012,** 24, (43), 5832-5836.
(35)     Perea-López, N.; Elías, A. L.; Berkdemir, A.; Castro-Beltran, A.; Gutiérrez, H. R.; Feng, S.; Lv, R.; Hayashi, T.; López-Urías, F.; Ghosh, S.; Muchharla, B.; Talapatra, S.; Terrones, H.; Terrones, M. *Adv. Funct. Mater.* **2013,** 23, (44), 5511-5517.
(36)     Hu, P.; Wen, Z.; Wang, L.; Tan, P.; Xiao, K. *ACS Nano* **2012,** 6, (7), 5988-5994.
(37)     Rose, A., *Concepts in photoconductivity and allied problems*. Interscience Publishers: 1963.
(38)     Bube, R. H., *Photoelectronic Properties of Semiconductors*. Cambridge University Press: 1992.
(39)     Konstantatos, G.; Clifford, J.; Levina, L.; Sargent, E. H. *Nat. Photonics* **2007,** 1, (9), 531-534.
(40)     Prins, F.; Buscema, M.; Seldenthuis, J. S.; Etaki, S.; Buchs, G.; Barkelid, M.; Zwiller, V.; Gao, Y.; Houtepen, A. J.; Siebbeles, L. D. A.; van der Zant, H. S. J. *Nano Lett.* **2012,** 12, (11), 5740-5743.
(41)     Castellanos-Gomez, A.; Wojtaszek, M.; Tombros, N.; Agraït, N.; van Wees, B. J.; Rubio-Bollinger, G. *Small* **2011,** 7, (17), 2491-2497.
(42)     Liu, F.; Shimotani, H.; Shang, H.; Kanagasekaran, T.; Zólyomi, V.; Drummond, N.; Fal'ko, V. I.; Tanigaki, K. *ACS Nano* **2013,** 8, (1), 752-760.

Fast and broadband photoresponse of a few-layer black phosphorous field-effect transistor.

*Michele Buscema\*, Dirk J. Groenendijk, Sofya I. Blanter, Gary A. Steele, Herre S.J. van der Zant and Andres Castellanos-Gomez\*.*

Kavli Institute of Nanoscience, Delft University of Technology, Lorentzweg 1, 2628 CJ Delft (The Netherlands).

Table of contents:

1) AFM micrographs of the studied devices

2) Effect of gate stress

3) Effect of illumination of different wavelengths to transfer curves for the studied devices

    a. Responsivity as a function of wavelength for the studied devices

    b. Responsivity as a function of power for the studied devices

4) Effect of illumination with 405 nm light

*AFM characterization of devices.*

In this section, we present the AFM chacterization of the studied devices. The measured height of the flakes ranges from 3 to 8 nm, corresponding to 6 to 16 layers of black phosphorus. The device showcased in the main text is represented in panel f.

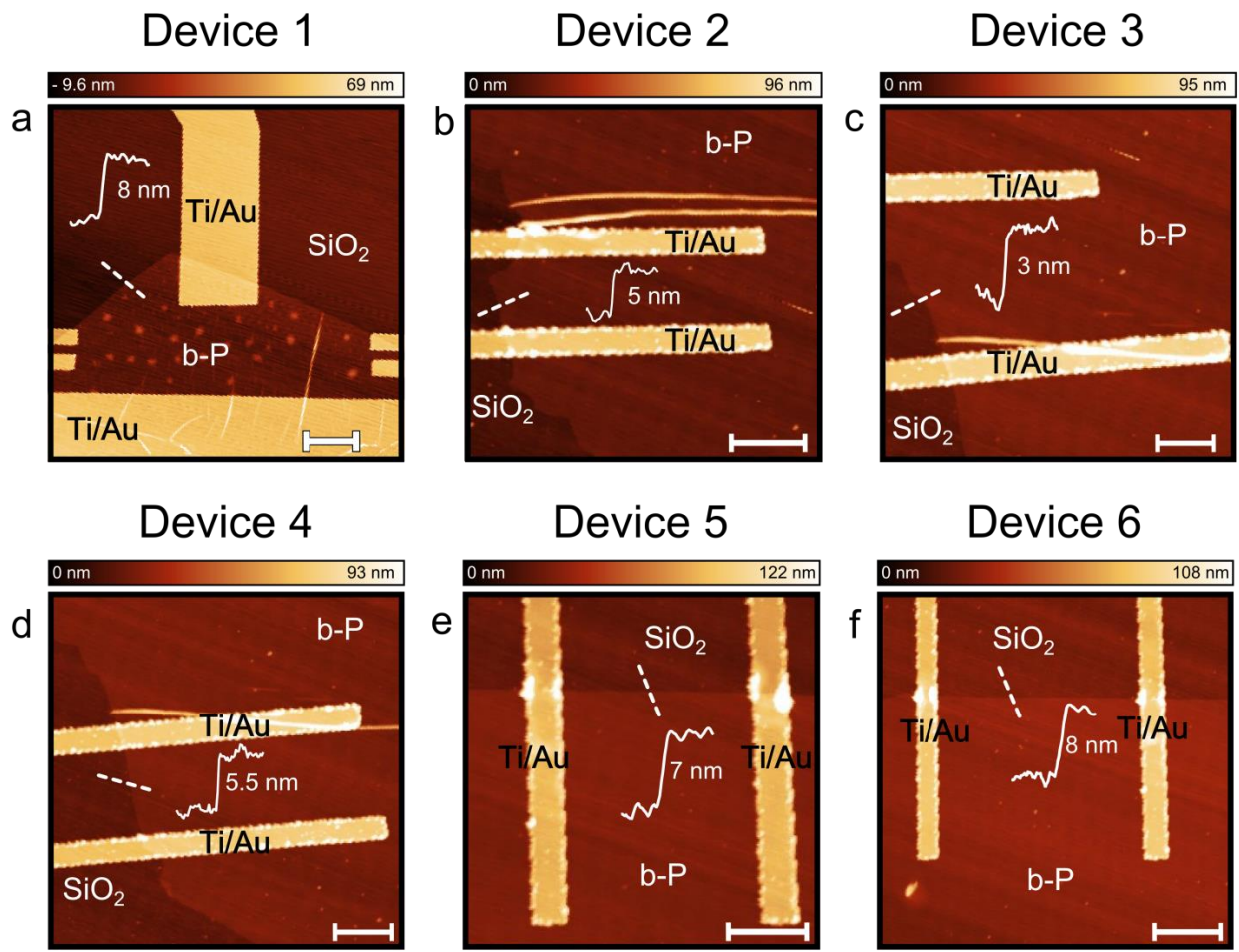

**Figure S1.** a to f) AFM characterization of the measured b-P devices. The scale bar is 3 μm in all panels.

*Effect of gate stress*

In this section we present the effect of the gate stress on the device performance. Figure S1(a) plots 50 consecutive transfer curves ($V_{ds}$ = 100 mV) and the inset shows a zoom in close to the threshold gate voltage. The effects of the consecutive gate cycles are evident: both the measured current and threshold voltage increase as more transfer curves are measured. This is a typical result of stress induced in the device by gate voltage cycles.[1,2] As further proof, we measure transfer curves before and after applying a constant gate voltage (Figure S1b). After 30 minutes of constant gate stress ($V_g$ = 30 V), the transfer characteristic (solid yellow line) of the device is shifted by ~ 10 V towards higher gate voltage values as compared to the starting point (solid blue line). The curve is rigidly shifted, so we measrue more current in the p-type configuration and a positive shift in the flatband condition. After letting the device relax ($V_g$ = 0 V) for 30 minutes, part of the induced shift is recovered (solid red line). This shift could be due to the filling of long-lived trap states induced by the gate or to charged species that are attracted and absorb at the surface of the b-P flake.[3]

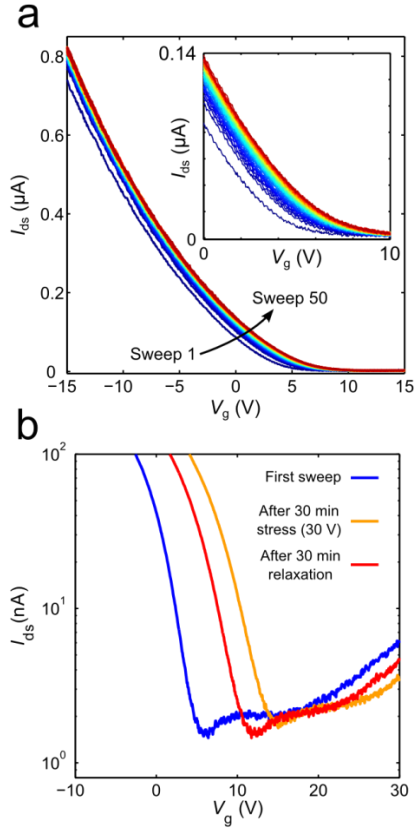

**Figure S2**. a) Series of consecutive transfer curves showing an increase in current from the first sweep to the last. The inset shows a zoom in close the flatband condition. b) Transfer curves showing the effect of gate bias stress. The reference measurement is plotted as blue solid line. The transfer curve measured after 30 minutes of constant gate stress (Vg = 30 V) is plotted as solid yellow line, and after 30 minutes of relaxation (Vg = 0 V) is plotted as solid red line.

*Effect of illumination with different wavelengths on the transfer curves for the studied devices*

In this section, we present the transfer characteristics in dark and under illumination for all the studied devices. Note that the indicated power is the total incident power not rescaled for the device area. All the devices show the same qualitative behavior: an increase of current under illumination. The shift in the flatband condition could be to a combination of gate stress and charging under light illumination (photonic gating). Device 6 is the device showcased in the main text.

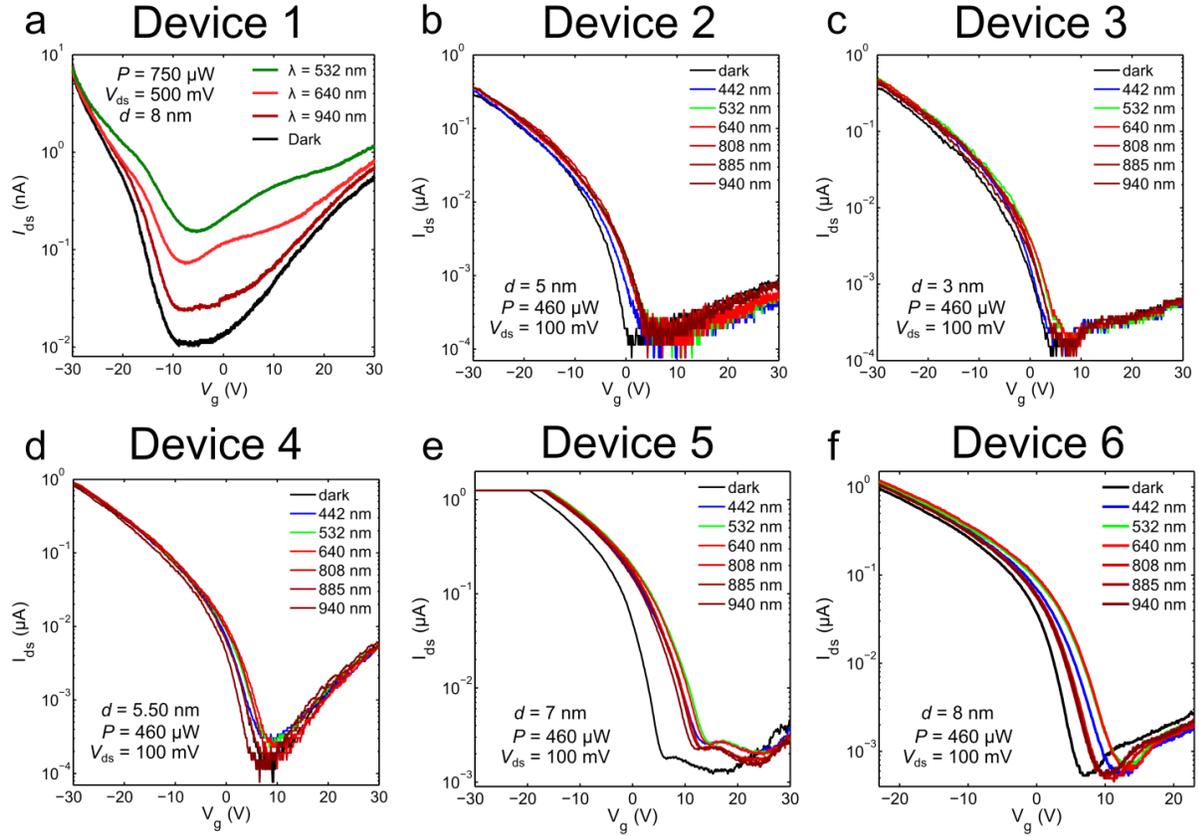

**Figure S3.** (a) to (f) Transfer curves of 6 different devices measured under illumination with different wavelengths, showing the same qualitative behavior. The indicated power is the total incident power, not rescaled for the device area.

*Responsivity as a function of wavelength for the studied devices*

The current level decreases for lower energy photons (i.e. increasing the photon wavelength), as can be seen from the extracted responsivity as a function of wavelength for all the measured devices (Figure S4). This could be explained by a reduced absorption of photons by the b-P flake as the excitation wavelength is increased.

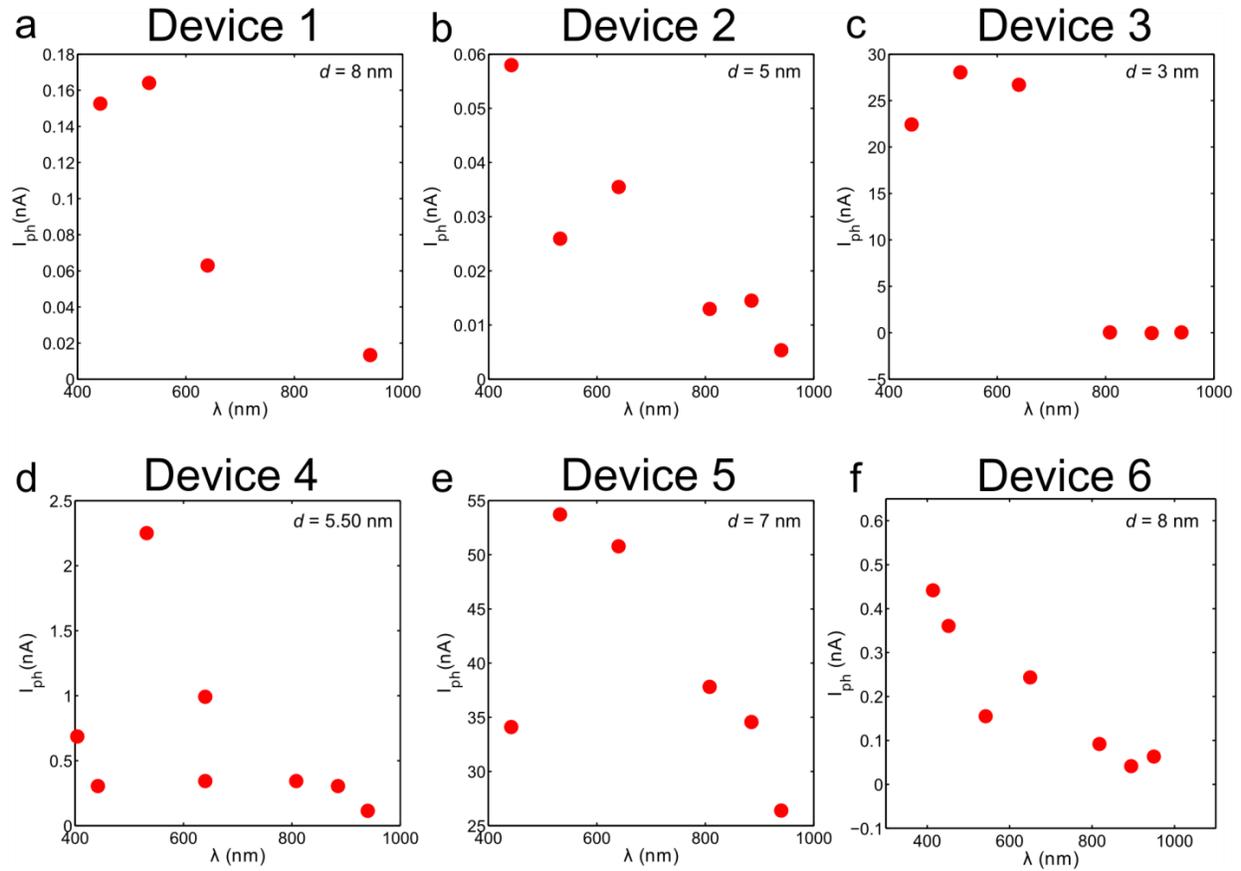

**Figure S4.** (a) to (f) Responsivity as a function of wavelength for 6 different devices. The responsivity is taken as the photocurrent at flatband condition of the reference dark measurement. Device 6 is the device showcased in the main text.

*Responsivity as a function of power for the studied devices*

By varying the excitation power at a fixed wavelength, the photocurrent increases sublinearly (Figure S5). This could be related to either filling of the trap states responsible for the photogeneration of carriers or increased recombination of (photogenerated) charge carriers while increasing the flux of incident photons. The photocurrent and the incident optical power are related by a power law relationship ($I_{ph} \propto P^{\alpha}$). The power law fits (dashed lines) to the experimental data (solid squares) have an exponent between 0.5 and 0.1, indicating that the saturation is due to a kinetics of the photogenerated carriers that involves both recombination-states and carrier-carrier interactions.

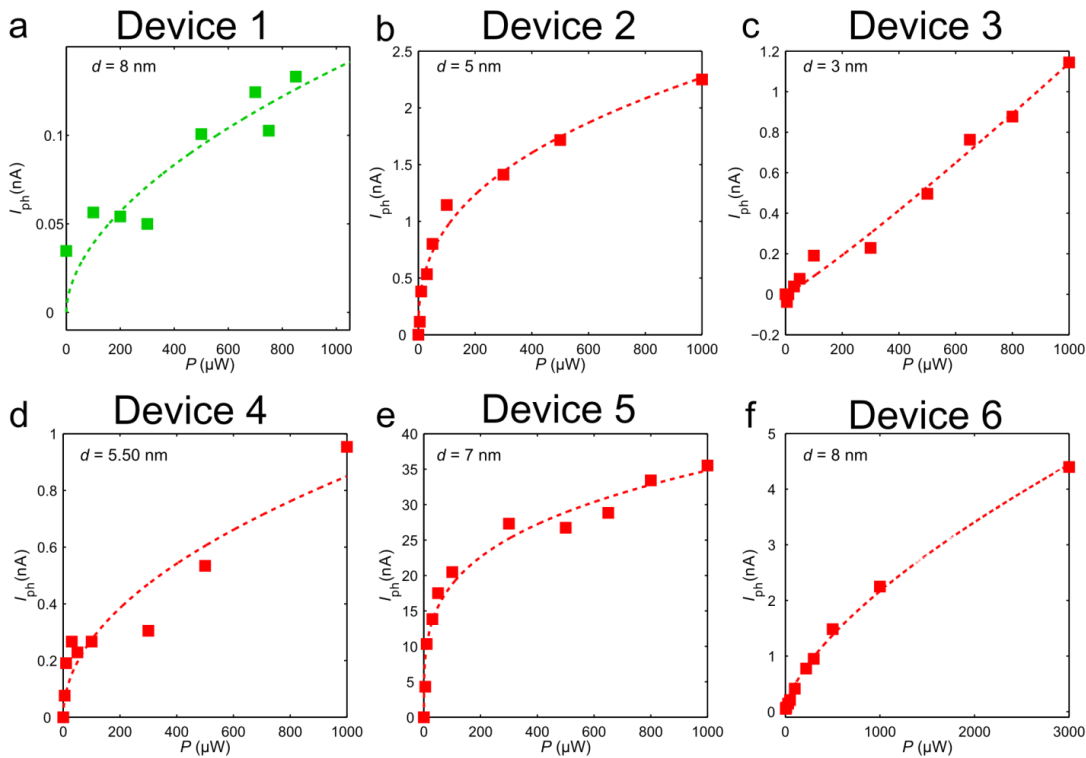

**Figure S5.** (a) to (f) Responsivity as a function of the total incident laser power (red squares) fitted by a power law for the studied devices (panel a, $\lambda = 532$ nm, panel b-f $\lambda = 640$ nm). The fitting exponent is in between 0.5 and 1 for all the plots. Device 6 is the device showcased in the main text.

*Effect of illumination with λ = 405 nm.*

In this section we present the data measured under illumination with λ = 405 nm. Under constant illumination with λ = 405 nm we measure a slow decrease of the average current level (Fig S6). Once the illumination is switched on, the current is seen to instantaneously increase and then drastically decrease until the illumination is removed. At that point, the current instantaneously decreases and then starts to slowly increase. This qualitative behavior is repeated in the following cycle of illumination. The instantaneous change in current upon illumination could be due to the fast extraction of photogenerated free carriers, which stops when illumination is turned OFF. The slow decrease (increase) of the current under continuous illumination (in dark) could be associated with photonic gating. The incident photons generate carriers that occupy long-lived trap states, acting as a local gate. Once the illumination is switched OFF, the filled trap states begin to empty, either by extraction or recombination with other carriers, and the local gating effect reduces, slowly increasing the current.

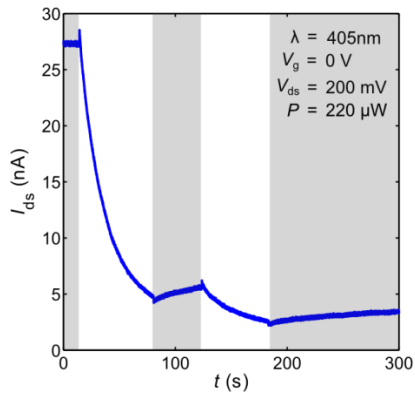

**Figure S6.** Measured current as a function of time under illumination with 405 nm laser and in dark (gray shaded areas).

Supplementary REFERENCES

(1)     Suresh, A.; Muth, J. F. *Applied Physics Letters* **2008,** 92, (3), -.

(2)     Gomes, H. L.; Stallinga, P.; Dinelli, F.; Murgia, M.; Biscarini, F.; de Leeuw, D. M.; Muck, T.; Geurts, J.; Molenkamp, L. W.; Wagner, V. *Applied Physics Letters* **2004,** 84, (16), 3184-3186.

(3)     Koenig, S. P.; Doganov, R. A.; Schmidt, H.; Neto, A. H. C.; Oezyilmaz, B. *arXiv:1402.5718* **2014**.